\RequirePackage{ifpdf}
\ifpdf 
\documentclass[pdftex]{sigma}
\else
\documentclass{sigma}
\fi

\newlength{\figwidth}
\newlength{\txtwidth}
\newlength{\parind}  
\def\miniabove{\vskip .5ex\noindent}
\def\miniparindent{\setlength{\parindent}{\parind}}
\def\minibelow{\vskip -\baselineskip \noindent}
\def\minijustify{\linebreak}

\def\d{{\rm d}}
\def\C{{\mathbb C}}
\def\R{{\mathbb R}}

\begin{document}

\allowdisplaybreaks
	
\renewcommand{\PaperNumber}{107}

\FirstPageHeading

\renewcommand{\thefootnote}{$\star$}

\ShortArticleName{Singular Potentials in Quantum Mechanics}

\ArticleName{Singular Potentials in Quantum Mechanics\\ and Ambiguity in
the Self-Adjoint Hamiltonian\footnote{This paper is a
contribution to the Proceedings of the 3-rd Microconference
``Analytic and Algebraic Me\-thods~III''. The full collection is
available at
\href{http://www.emis.de/journals/SIGMA/Prague2007.html}{http://www.emis.de/journals/SIGMA/Prague2007.html}}}

\Author{Tam\'as F\"UL\"OP}

\AuthorNameForHeading{T. F\"ul\"op}

\Address{Montavid Research Group, Budapest,
Soroks\'ari~\'ut~38-40, 1095, Hungary}
\Email{\href{mailto:tamas.fulop@gmail.com}{tamas.fulop@gmail.com}}

\ArticleDates{Received August 07, 2007, in f\/inal form November 08, 2007; Published online November 16, 2007}

\Abstract{For a class of singular potentials, including the Coulomb
potential (in three and less dimensions) and $V(x) = g/x^2$ with the
coef\/f\/icient $g$ in a certain range ($x$ being a space coordinate in one
or more dimensions), the corresponding Schr\"{o}dinger operator is not
automatically self-adjoint on its natural domain. Such operators admit
more than one self-adjoint domain, and the spectrum and all physical
consequences depend seriously on the self-adjoint version chosen. The
article discusses how the self-adjoint domains can be identif\/ied in terms
of a boundary condition for the asymptotic behaviour of the wave
functions around the singularity, and what physical dif\/ferences emerge
for dif\/ferent self-adjoint versions of the Hamiltonian. The paper reviews
and interprets known results, with the intention to provide a practical
guide for all those interested in how to approach these ambiguous
situations.}

\Keywords{quantum mechanics; singular potential; self-adjointness;
boundary condition}

\Classification{81Q10}

\renewcommand{\thefootnote}{\arabic{footnote}}
\setcounter{footnote}{0}

\section{Introduction}

Let us consider a quantum mechanical Schr\"{o}dinger Hamiltonian
  \begin{equation}  \label{eq:aaa}
  H = - \frac{\hbar^2}{2m} \bigtriangleup + V,
  \end{equation}
where the real potential $V$ is singular, for example,
$ V(r) \sim \frac{1}{r} $ or $ V(r) \sim \frac{1}{r^2} $. To keep the
discussion technically simple, let our conf\/iguration space be a one
dimensional interval or half line or line. Remarkably, the principles
and methods to come will be valid for higher dimensional conf\/iguration
spaces, too, it is only the amount of technicalities that increases with
the dimensions. It is also important to note that the half line case
plays a direct role in a number of higher dimensional problems as well.
For instance, in a higher dimensional setting with a central symmetric
potential that is divergent at the centre, one can perform a
separation of variables in a spherical coordinate system, and the
ambiguity that will be our topic here appears in the radial part of the
system, which is actually a half line problem.

Now, in a straightforward manner, let us innocently search for all the
eigenfunctions of this dif\/ferential operator. Then, depending on our
``luck", we may encounter some surprising
dif\/f\/iculties\footnote{Naturally, this will not be a question of ``luck",
and we will see soon the mathematical criterion that tells that which
potentials cause these dif\/f\/iculties.}. Namely, we may f\/ind that, actually,
there are ``too many" eigenfunctions, in the sense that they are not
mutually orthogonal, are not linearly independent, and form an
overcomplete system rather than a basis in the Hilbert space of square
integrable functions. Moreover, we can observe that there are so many
eigenfunctions that even the eigenvalues are not restricted to real
values but any nonreal $ \lambda \in \C \setminus \R$ also proves to be
the eigenvalue of some (square integrable) eigensolution of the
eigenvalue problem. Apparently, the self-adjointness of our, naively
self-adjoint, Hamiltonian is seriously challenged. This can also be seen
from that, for generic wave functions $\psi$ and $\chi$, applying
integration by parts,
  \begin{equation}  \label{eq:aab}
  ( H \psi, \chi ) - ( \psi, H \chi ) = \hbox{ surface\ terms } \ne 0
  \end{equation}
will be observable.

Now, if we don't have self-adjointness then we have no spectral theorem,
no physical interpretation, and no unitary time evolution. Therefore, we
are heavily motivated to restore self-adjointness, if only possible.

This last observation \eqref{eq:aab} may suggest us that the problem is
created at the boundaries, especially at the location of singularity. We
can try to cure the situation by requiring that all wave functions should
vanish or decrease fast enough as we approach the singularity so that the
surface terms tend to zero. Unfortunately, we will f\/ind that this way we
lose all the eigenfunctions. Hence, this requirement cannot help in
solving the problem. It seems that we need to keep some of the
eigenfunctions and to dispose the rest of them.

Actually, the usual reaction to the problem  --  as many textbooks treat
the three dimensional Coulomb problem, for example  --  is to declare that
not all eigenfunctions are ``acceptable". Suddenly, some extra ad hoc
requirement is invented, which does not follow from the axioms/(general
principles) of quantum theory but is introduced on the f\/ly. Unfortunately,
it is beyond the scope of this discussion to analyze those various
conditions in detail. To give only a summary of the result of such a
careful and honest analysis, one can reveal that
\begin{itemize}\itemsep=0pt
\item
those requirements are physically questionable,
\item
they have mathematically limited availability,
\item
dif\/ferent conditions can lead to dif\/ferent results,
\item
many consistent and valuable quantum mechanical models are lost by those
requirements.
\end{itemize}
The situation is well illustrated, for example, by the paper
\cite{ref:GC}, which, for the attractive Coulomb potential in one
dimension, gives a critical overview of various existing treatments and
their various dif\/ferent results -- ands adds another new approach, which
is also questionable both mathematically and physically.

However, we can follow another philosophy as well. Instead of the urge to
choose, we can accept that the potential itself does not f\/ix the model
uniquely. When an application forces us to f\/ix the ambiguity, some
additional physical information will be needed. That information can come
from experimental measurement or, if available, from some additional
theoretical knowledge about the concrete case.

Therefore, for these potentials, let us search for all the quantum
mechanically allowed cases (self-adjoint Hamiltonians) related to our
initial Hamiltonian. Then, when in a concrete problem we need to choose
one,
\begin{itemize}\itemsep=0pt
\item
knowing what possibilities exist may help to choose,
\item
if we have some additional information then it will be easier to utilize it,
\item
if measurement is needed to decide then we know what to measure and what
to f\/it to the experimental data.
\end{itemize}

To carry out f\/inding all the cases, it is advisable to review f\/irst what
mathematics knows and tells about self-adjointness and about the possible
ambiguity in it.

\section{Self-adjoint extensions of symmetric operators}  \label{sec:math}

We will make use of the following def\/initions and
theorems\footnote{All mathematical ingredients quoted here and hereafter
are taken from the sources
\cite{ref:RSI,ref:RSII,ref:Richtmyer,ref:AkhGl,ref:GorbGorb}.}
(the validity
of which is, nevertheless, not restricted to dif\/ferential operators).

Let $A$ be a linear operator that is def\/ined on a dense
subset ${\cal D}(A)$ of a separable Hilbert space~${\cal H}$.

{\it The adjoint} $A^+$ of $A$ is def\/ined on those vectors
$\chi \in {\cal H}$ for which there exists a
$\tilde{\chi} \in {\cal H}$ such that
  \begin{equation*}
  (A\psi, \chi) = (\psi, \tilde{\chi}) \qquad \mbox{for} \quad
  \forall  \, \psi \in {\cal D}(A) ,
  \end{equation*}
and $A^+$ is def\/ined on such a $\chi$ as $ A^+ \chi := \tilde{\chi} $.

$A$ is called {\it symmetric} if
  \begin{equation*}
  (A\psi, \chi) = (\psi, A\chi) \qquad \mbox{for} \quad
  \forall \, \psi, \chi \in {\cal D}(A) .
  \end{equation*}
The adjoint of a symmetric $A$ is always an extension of it (i.e.,
${\cal D}(A)$ is a subset of ${\cal D}(A^+)$, and $A$ and $A^+$ act the
same way on ${\cal D}(A)$).

$A$ is {\it self-adjoint} if $A=A^+$. (Which includes that their domains
coincide.)
$A$ is {\it essentially self-adjoint} if it admits a unique self-adjoint
extension.

Remarkably, not the symmetric but only the self-adjoint operators are
those for which the spectral theorem holds, and which are in a one-to-one
correspondence with the strongly continuous one-parameter
unitary groups $U(t)$ on ${\cal H}$ (to any $U(t)$, there is a unique
self-adjoint $A$ such that $U(t) = e^{-iAt}$). Both these properties play
an important role in the physical interpretation of quantum mechanics.

Next, from now on, let $A$ be a symmetric operator that is {\it closed}
(for symmetric operators, this simply means the requirement $A^+{}^+ =
A$; any symmetric operator admits a closure~-- a~minimal closed
extension -- which is actually nothing but its double adjoint). Let us
also f\/ix an arbitrary nonreal number $\lambda$.

The {\it deficiency subspaces} ${\cal E}_\lambda$ and
${\cal E}_{\lambda^*}$ are def\/ined as the eigensubspace of $A^+$ with
respect to the eigenvalue $\lambda$ and the complex conjugate eigenvalue
$\lambda^*$, respectively. The {\it deficiency indices}~$n_\lambda$ and~$n_{\lambda^*}$ are the dimension of ${\cal E}_\lambda$, resp.~${\cal E}_{\lambda^*}$. $A$ is self-adjoint if and only if $ n_\lambda =
n_{\lambda^*} = 0 $.

$A$ admits self-adjoint extensions if and only if its def\/iciency indices
are equal\footnote{For our dif\/ferential operator Hamiltonian
\eqref{eq:aaa} with a real potential, this will always be the case,
since such an~operator is invariant under complex conjugation.},
$ n_\lambda = n_{\lambda^*} =: n $. The self-adjoint extensions are in a
one-to-one correspondence with the unitary maps from ${\cal E}_\lambda$ to
${\cal E}_{\lambda^*}$. Each unitary
$ U_{\rm N} : {\cal E}_\lambda \to {\cal E}_{\lambda^*} $
characterizes a self-adjoint extension $A_{U_{\rm N}}$ as the restriction
of $A^+$ to the domain
  \begin{equation}  \label{eq:aae}
  {\cal D}(A_{U_{\rm N}}) = \{ \psi_0 + \psi_\lambda + U_{\rm N}
  \psi_\lambda \; | \; \psi_0 \in {\cal D}(A), \;
  \psi_\lambda \in {\cal E}_\lambda \} .
  \end{equation}

The unitary maps $U_{\rm N}$ act on an $n$ dimensional space so they can
be param\-etr\-ized by $n^2$ real parameters\footnote{For our dif\/ferential
operator Hamiltonian in one dimension, $n$ is f\/inite (unless the
potential admits inf\/initely many singularities). In higher dimensions,
$n$ can be $\infty$, which happens when the singular locations are not
f\/initely many points but lie, say, along a line.}.

After the characterization \eqref{eq:aae} given by von Neumann, let us
see a more recent alternative description of the possible self-adjoint
extensions, the so-called boundary value space approach:

If there exists an (auxiliary) Hilbert space ${\cal H}_{\rm b}$
(necessarily $n$ dimensional) and two linear maps
$ \Gamma^{}_1, \Gamma^{}_2 : {\cal D}(A^+) \to {\cal H}_{\rm b} $
such that, for $ \forall \psi, \chi \in {\cal D}(A^+) $,
  \begin{equation}  \label{eq:aaf}
  (A^+\psi, \chi) - (\psi, A^+\chi) = (\Gamma^{}_1 \psi, \Gamma^{}_2
  \chi)^{}_{\rm b} - (\Gamma^{}_2 \psi, \Gamma^{}_1 \chi)^{}_{\rm b} ,
  \end{equation}
and to any two $\Psi^{}_1$, $\Psi^{}_2 \in {\cal H}_{\rm b}$
there exists a $ \psi \in {\cal D}(A^+) $ satisfying
  \begin{equation}  \label{eq:aag}
  \Gamma^{}_1 \psi = \Psi^{}_1 , \qquad \Gamma^{}_2 \psi = \Psi^{}_2 ,
  \end{equation}
then there is a one-to-one correspondence between the unitary maps
$ U \in {\cal U}({\cal H}_{\rm b}) $ and the self-adjoint extensions of
$A$, where a $U$ describes the domain of the corresponding $A_U$ as
  \begin{equation*}
  {\cal D}(A_U) = \{ \psi \in {\cal D}(A^+) \; | \;
  (U - \mathbf{1}_{\rm b}) \Gamma^{}_1 \psi +
  i (U + \mathbf{1}_{\rm b}) \Gamma^{}_2 \psi = 0 \} .
  \end{equation*}

Heuristically, the second condition, related to \eqref{eq:aag}, says
that the auxiliary boundary Hilbert space ${\cal H}_{\rm b}$ must be a
smallest one among those fulf\/illing the f\/irst condition,
\eqref{eq:aaf}. If an ${\cal H}_{\rm b}$ is suitable for \eqref{eq:aaf}
then some appropriate trivial ``enlargement", extension of it can also be
suitable, e.g., by orthogonally adding
some other Hilbert space to it, so this second condition is to ensure the
ef\/f\/iciency of the description by removing the redundancy.

One can f\/ind that the former characterization of self-adjoint domains can
be considered as a special case of the latter, with $ {\cal H}_{\rm b} =
{\cal E}_\lambda $. Another note to make is that, for a f\/ixed $ {\cal
H}_{\rm b} $, the choice of appropriate $\Gamma$s is not unique. Hence,
it is not unique that which $U$ provides which self-adjoint version.
Therefore, one should not -- at least at the general level -- attribute any
distinguished meaning to that, having one choice of ${\cal H}_{\rm b}$,
$\Gamma^{}_1$ and $\Gamma^{}_2$, which self-adjoint version is indexed by
which unitary operator. (See more on it later.)
A third remark is that, although the boundary value
space approach may appear a rather abstract equipment at f\/irst sight, in
applications we can f\/ind it simple, friendly and handy. Actually,
historically it has been worked out to directly suit the special cases of
dif\/ferential operators where the ``boundary values" \eqref{eq:aag} are
indeed the limiting values of the wave function and its derivative at the
boundary of the conf\/iguration space  --  or appropriate combinations of
them. Soon we will see examples that show how practical this approach is
for dif\/ferential operators.

\section{Finding the self-adjoint Hamiltonians}  \label{sec:applmath}

\setlength{\figwidth}{.35\textwidth}
\setlength{\txtwidth}{\textwidth} \addtolength{\txtwidth}{-1.2\figwidth}
\begin{minipage}[b]{\txtwidth}
Applying the content of the previous section to our dif\/ferential operator
of the type $ - \frac{\hbar^2}{2m} \bigtriangleup + V $, we f\/irst need to
specify an initial domain on which it is symmetric. We can choose, for
example, those wave functions~$\psi $ which admit a continuous second
derivative, vanish in a neighbourhood of any singularity of the
potential and near any f\/inite endpoints of our conf\/iguration
space (if applicable), have compact support (so that
\minijustify
\end{minipage}
\hfill
\begin{minipage}[b]{\figwidth}
\raisebox{.14\textwidth}{\resizebox{\textwidth}{!}
{\includegraphics{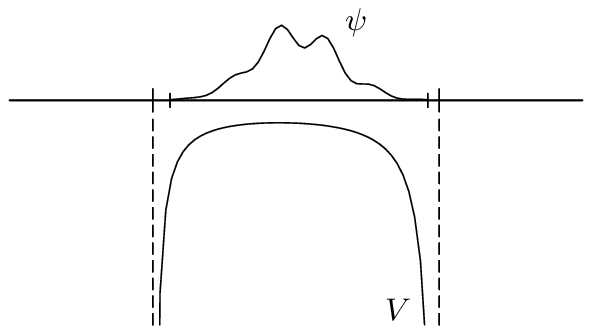}}}
\end{minipage}
\minibelow
they also vanish ``in a
neighbourhood of inf\/inity", for the case when the conf\/iguration space has
$+\infty$ and/or $-\infty$ as ``endpoint"), and $\psi$, $H \psi$ are
square integrable. See the f\/igure for an illustration. Such a domain is
dense in $L^2$.

The closure of this symmetric operator will have a bit generalized
domain, where the smoothness property is weaker ($\psi$ and its
derivative must be absolutely continuous), and the wave functions do not need
to vanish identically around any ``problematic" location but only to
decrease fast enough so that the limiting surface terms of
\eqref{eq:aab} emerging at the boundaries and singularities will be zero
for any pairs $\psi$, $\chi$. At last, the adjoint of this closed
operator will have the further generalized domain in which there is no
restriction on the behaviour around the ``problematic" locations.

Now, to specify the self-adjoint domains, which lie in between the
symmetric domain and the adjoint domain, let us use the boundary
value space description because that requires the least calculational
ef\/forts, and because there are known recipes how to f\/ind suitable
candidates for the needed ingredients ${\cal H}_{\rm b}$,
$\Gamma^{}_1$, $\Gamma^{}_2$. In what follows, we will see some
instructive examples how the method works in practice. The f\/irst two
examples are chosen to be simple to make the essence of the procedure
apparent. In the meantime, they will already exhibit many of the
general physical properties that typically arise when one has a
singularity-induced or boundary-induced ambiguity in self-adjointness.

\section{Free particle on a half line}  \label{sec:halffree}

\setlength{\figwidth}{.26\textwidth}
\setlength{\txtwidth}{\textwidth} \addtolength{\txtwidth}{-1.3\figwidth}
\miniabove
\begin{minipage}[b]{\txtwidth}
As the f\/irst example, let us consider a free particle moving on a half
line, which we wish to be bordered by a perfectly ref\/lecting boundary to
ensure the conservation of probability and, correspondingly, a
self-adjoint domain for the Hamiltonian. (See the f\/igure for an
illustration and for the notations.)
\minijustify
\end{minipage}
\hfill
\begin{minipage}[b]{\figwidth}
\raisebox{.2\textwidth}{\resizebox{\textwidth}{!}
{\includegraphics{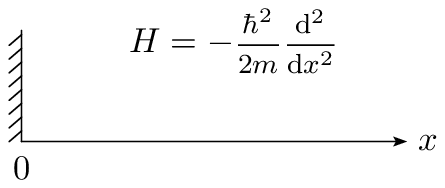}}}
\end{minipage}
\minibelow
The most well-known way to reach this is to
impose the Dirichlet boundary condition $\psi(0) = 0$. However, other
boundary conditions also provide self-adjoint possibilities, and
physically it is a~natural idea that there can be various dif\/ferent types
of ref\/lecting walls\footnote{We can imagine, for example, a long and
thin nanowire, with a small piece of some material attached to its ends.
Then, near one such end, quantum ef\/fects in the range of wavelengths that
are much larger then the width of the wire as well as the size of this
attached blob but smaller then the length of the wire may be well
described by one of the ``free particle on the half line" self-adjoint
Hamiltonians, presumably dif\/ferent ones for dif\/ferent materials
attached.}. Therefore, exploring all the self-adjoint versions means to
f\/ind the variety of perfectly ref\/lecting walls that can be accounted for
in the scope of this quantum mechanical model.

Starting with a symmetric domain and considering the corresponding adjoint
domain as discussed in the previous section, integration by parts allows
us to write, for the adjoint operator,
  \begin{equation*}
  (H^+\psi, \chi) - (\psi, H^+\chi)
  = - \frac{\hbar^2}{2m} (\psi^* \chi' - {\psi^*}'\chi)(+0)
  = - \frac{\hbar^2}{2m} W[\psi^*, \chi](+0) ,
  \end{equation*}
where $W[\cdot, \cdot]$ denotes the Wronskian. In principle, we could
expect a surface term related to $x \to \infty$ as well (since it is not
immediately clear whether the square integrability of $\psi$ and
$H^+ \psi$ plus the degree of smoothness $\psi$ has provides a fast
enough decrease towards inf\/inity to send that term to zero): it will be
explained in Section~\ref{sec:singpot}, at a general level, why it is
absent indeed.

Comparing this formula with \eqref{eq:aaf} shows that we can comply with
the boundary value space approach with the choices
  \begin{equation*}
  {\cal H}_{\rm b} = \C, \qquad \Gamma^{}_1 \psi = \psi(0),
  \qquad \Gamma^{}_2 \psi = L^{}_0 \psi'(0).
  \end{equation*}
Here, we have introduced an auxiliary nonzero real length $L^{}_0$, which
does not play any principal role in the question of self-adjointness but
is needed merely on dimensional grounds. Namely, the formalism requires
$\Gamma^{}_1$ and $\Gamma^{}_2$ to have the same dimensions while $\psi$
and $\psi'$ dif\/fer in a dimension of length\footnote{For a full theory
of how physical dimensions can be formulated and treated
math\-ematically -- via one dimensional vector spaces and their tensorial
products, quotients and powers -- see \cite{ref:Matolcsi}.}. Also, we
have dropped the factor $-\frac{\hbar^2}{2m}$ (that is, \eqref{eq:aaf}
has actually been fulf\/illed for $A := \frac{2m L^{}_0}{\hbar^2} H$).

The other, ``ef\/f\/iciency", condition needed for the boundary value space
approach is also not hard to check. This point we are actually going to
discuss later in Section~\ref{sec:singpot}, again at a general level.

Having fulf\/illed the requirements of the boundary value space method, we
are allowed to harvest the fruit: the possible self-adjoint domains form
a one-parameter family, indexed by $U \in {\cal U}(1)$, in a parametrized
form, $U = e^{i \vartheta}$, $\vartheta \in [0, 2\pi)$, via the boundary
condition
  \begin{equation*}
  \big( e^{i \vartheta} - 1 \big) \psi(0) +
  i \big( e^{i \vartheta} + 1 \big) L^{}_0 \psi'(0) = 0 .
  \end{equation*}
We can rewrite this condition in the more simplif\/ied form
  \begin{equation*}
  \psi(0) + L \psi'(0) = 0 ,  \qquad  L = L^{}_0 \, \hbox{cot}
\frac{\vartheta}{2} \in (-\infty, \infty) \cup \{\infty\}
  \end{equation*}
as well. Hence, the quantum mechanically allowed ref\/lecting walls are
characterized by an arbitrary real length parameter. The Dirichlet case
corresponds to $L=0$, the Neumann boundary condition $\psi'(0) = 0$ is
the case $L = \infty$, and the remaining cases, containing a nontrivial
combination of the wave function and its derivative in the boundary
condition, are often called the Robin boundary conditions.

It is instructive to see how the spectral properties, and correspondingly
all physical properties, depend on the boundary parameter $L$. Omitting
technical details, solving the eigenvalue problem for a given boundary
condition yields \cite{ref:d2,ref:wall}
  \begin{align*}
  E > 0:  & \quad  \varphi_k(x) =
  \frac{1}{\sqrt{2\pi}} ( e^{-ikx} - e^{2i\delta_k} e^{ikx} ) ,
  \qquad  e^{2i\delta_k} = \frac{1 - ikL}{1 + ikL} ,
  \nonumber \\
  E < 0:  &  \quad  \varphi_{\rm bound}(x) =
  \sqrt{\frac{2}{L}} e^{-x/L} ,  \qquad  E_{\rm bound} =
  - \frac{\hbar^2}{2mL^2}  \quad  (\hbox{only if } L > 0) .
  \end{align*}
Therefore, for example the existence and energy of a boundary bound state
is heavily $L$-dependent. The $L$-dependence of the scattering phase shift
can also be visualized by, e.g., the
\linebreak
\setlength{\figwidth}{.33\textwidth}
\setlength{\txtwidth}{\textwidth} \addtolength{\txtwidth}{-1.3\figwidth}
\vskip -1.75ex\noindent
\begin{minipage}[b]{\txtwidth}
time delay, which is the dif\/ference
between the time when the peak of an incoming wave packet reaches the
wall and the time when the peak of the ref\/lected packet leaves the wall.
For an incoming packet concentrated in wave number around the value $-k$,
the result is \cite{ref:d2,ref:wall}
  \begin{equation*}
  \tau = \frac{-2mL}{\hbar k (1 + k^2 L^2)} ,
  \end{equation*}
\minijustify
\end{minipage}
\hfill
\begin{minipage}[b]{\figwidth}
\raisebox{.21\textwidth}{\resizebox{\textwidth}{!}
{\includegraphics{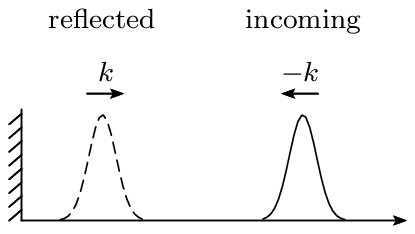}}}
\end{minipage}
\minibelow
which seriously depends on $L$ (even its sign is decided by
the sign of $L$). For further $L$-related ef\/fects and aspects, the Reader
is asked to consult \cite{ref:d2,ref:wall}.

\section{Free particle on a line with a point interaction}
 \label{sec:linefree}

\setlength{\figwidth}{.33\textwidth}
\setlength{\txtwidth}{\textwidth} \addtolength{\txtwidth}{-1.3\figwidth}
\miniabove
\begin{minipage}[b]{\txtwidth}
In the next example, the particle can move on a~line freely, except at
one point where some object or disturbance or short-range potential-like
ef\/fect resides and can perform some nontrivial action on the particle.
\minijustify
\end{minipage}
\hfill
\begin{minipage}[b]{\figwidth}
\raisebox{.23\textwidth}{\resizebox{\textwidth}{!}
{\includegraphics{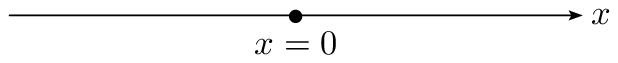}}}
\end{minipage}
\minibelow
 In
this setting, the widely known example is the Dirac delta potential but
the full family of possibilities again proves to be larger, and can be
explored again via f\/inding the possible self-adjoint versions of the
initial Hamiltonian.

We can proceed similarly to the half line case, writing
  \begin{equation*}
  (H^+\psi, \chi) - (\psi, H^+\chi) =
  - \frac{\hbar^2}{2m} \left\{ (\psi^* \chi' - {\psi^*}'\chi)(+0)
  - (\psi^* \chi' - {\psi^*}'\chi)(-0) \right\} ,
  \end{equation*}
and f\/inding agreement with \eqref{eq:aaf} through
  \begin{equation}  \label{eq:aaq}
  {\cal H}_{\rm b} = \C^2 ,  \qquad
  \Gamma^{}_1 \psi =
  \begin{pmatrix} \psi(+0) \cr \psi(-0) \end{pmatrix} ,  \qquad
  \Gamma^{}_2 \psi =
  L_0 \begin{pmatrix} \psi'(+0) \cr -\psi'(-0) \end{pmatrix} .
  \end{equation}
Correspondingly, the family of self-adjoint possibilities can be indexed
by $U \in {\cal U}(2)$, via
  \begin{equation*}
  \big( U - \mathbf{1}^{}_{\C_{}^2} \big) \Gamma^{}_1 \psi +
  i \big( U + \mathbf{1}^{}_{\C_{}^2} \big) \Gamma^{}_2 \psi = 0 .
  \end{equation*}
Similarly to $L$ of the half line case, here, the four real parameters
parametrizing the $U$s can be chosen as two length scales $L^{}_+$,
$L^{}_-$ and two angles. Roughly speaking, the reason for two length
parameters is that, now, the singular point has two sides. In parallel,
one of the angles expresses the mixing between the left and the right
side, and the other characterizes the strength of a~short-range vector
potential and thus causes a phase jump in the wave function at the
singular point\footnote{When replacing the line with a circle, this
fourth parameter corresponds to the magnetic f\/lux driven through the
circle.}. In the case of zero mixing, which happens when the matrix $U$ is
diagonal, there is no probability f\/low crossing the singularity, the two
half lines physically decouple, and the system becomes a sum of two
independent subsystems, two half line systems indeed (one with $L =
L^{}_+$ and the other with $L = L^{}_-$).

Just as the sign of $L$ decides the number of bound states in the half
line model, here the signs of $L^{}_+$ and $L^{}_-$
govern the number of bound states, which can thus be 0, 1 or 2. The
scattering properties also depend on the boundary parameters, causing
time delays, and making the point object to act as a low-pass or
high-pass f\/ilter depending on the values of the parameters. See
\cite{ref:d2,ref:lineduallet,ref:linefull,ref:mobius} for further
physical ef\/fects caused by the properties of the pointlike object.

\section{Singular potentials}  \label{sec:singpot}

Now, we enter the situation having a potential that possesses not an
abrupt (``hard") singularity like a Dirac delta but a continuously
developing one like for the Coulomb potential (``soft" singularity). The
complication with respect to the abrupt cases is that, here, the wave
functions in the adjoint domain and their derivative are typically
diverging when approaching the location of singularity. Therefore, we
cannot choose such limits as boundary values. However, there exist some
f\/inite numbers hidden in the asymptotics of the wave functions, which can
be extracted as follows. Let us f\/irst restrict ourselves to one side of
the singularity.

One can start with observing\footnote{For the mathematical results quoted
in this Section, see, e.g.,
\cite{ref:RSII,ref:Richtmyer,ref:AkhGl}.} that,
for any $\psi$, $\chi$ in the adjoint domain, $W[\psi^*, \chi]$ has a
f\/inite limit when approaching the singularity.
Next, f\/ixing an arbitrary real value $E_0$, let us consider two real
eigenfunctions $\varphi^{(1)}$, $\varphi^{(2)}$ of our Hamiltonian as a dif\/ferential
operator (i.e., omitting the requirements related to square
integrability) for the eigenvalue $E_0$, two such solutions that satisfy
$W[\varphi^{(1)}, \varphi^{(2)}] = 1$. We will call them {\it reference modes}.
Two possibilities exist:
\begin{enumerate}\itemsep=0pt
\item
Both reference modes are square integrable in a (one-sided) neighbourhood
of the location of the singularity  --  this is called in the
mathematical literature the {\it limit-circle case}\footnote{The name
has nothing to do with the geometry of our conf\/iguration space but is
because of some historical and technical reasons.}
(example: $ \frac{\hbar^2}{2m}
\left( - \frac{\d^2}{\d x^2} + \frac{5/16}{x^2} \right) $
on $(0, \infty)$, $E_0 = 0$,
$\varphi^{(1)}(x) = x^{5/4}$, $\varphi^{(2)}(x) = - \frac{2}{3} x^{-1/4}$),
\item
At most, only a specif\/ic linear combination of
them\footnote{Up to an overall complex
factor, of course.} is square integrable around the singularity  --
this is called the {\it limit-point case}
(example: $ \frac{\hbar^2}{2m}
\left( - \frac{\d^2}{\d x^2} + \frac{21/16}{x^2} \right)$,
on $(0, \infty)$, $E_0 = 0$,
$\varphi^{(1)}(x) = x^{7/4}$, $\varphi^{(2)}(x) = - \frac{2}{5} x^{-3/4}$).
\end{enumerate}

It turns out that, if our problem is in the limit-point case then,
for any two wave functions~$\psi$,~$\chi$ in the adjoint domain, $
W[\psi^*, \chi] $ tends to zero at the singularity. Hence, a limit-point
singularity does not create nonzero surface terms in \eqref{eq:aab} and,
consequently, does not induce an ambiguity in self-adjointness. As a
special case, this is the reason why, in the ``free particle on a half
line" system, we did not need to worry about the surface terms towards
inf\/inity. Note that an inf\/inite ``endpoint" is always to be treated as a
singular point, which can be seen, for example, from that the change of
variable $x \to \frac{1}{x}$ brings it into f\/inite (to zero), and via the
accompanying unitary transformation $ \tilde{\psi}(x) := \frac{1}{x}
\psi(\frac{1}{x}) $ our symmetric dif\/ferential operator
is mapped into a one that is singular at zero (e.g.,
$ - \frac{\hbar^2}{2m} \frac{\d^2}{\d x^2} $ is mapped to
$ - \frac{\hbar^2}{2m} \left[ \frac{\d}{\d x} \left( x^4 \frac{\d}{\d x}
\right) + 2 x^2 \right] $, and see the last paragraph of this Section).

In the limit-circle case where the surface terms do not vanish, what can
be shown is that any $\psi$ in the adjoint domain is asymptotically
similar to an appropriate linear combination of the reference modes, in
the sense that
  \begin{equation*}
  \psi(x) = \big[ c^{(1)} + \eta^{(1)}(x) \big] \varphi^{(1)}(x) +
  \big[ c^{(2)} + \eta^{(2)}(x) \big] \varphi^{(2)}(x) ,
  \end{equation*}
where $\eta^{(1)}(x)$, $\eta^{(2)}(x)$ vanish towards the singularity,
and the {\it limit numbers} $c^{(1)}$, $c^{(2)}$ can be read of\/f as the
(always f\/inite) limiting values of $- W[\varphi^{(2)}, \psi]$, respectively
$W[\varphi^{(1)}, \psi]$, at the singularity. Furthermore, this asymptotic
similarity proves to be strong enough for that, when we want to calculate
the limit of a $W[\psi^*, \chi]$ at the singularity, we can replace
$\psi(x)$ in it with $ c^{(1)} \varphi^{(1)}(x) + c^{(2)} \varphi^{(2)}(x) $ (and
$\chi$ can also be replaced with its similar approximation).

Adding
  \begin{equation*}
  W[\psi^*, \chi] = W[\varphi^{(1)}, \psi]^* W[\varphi^{(2)}, \chi] -
  W[\varphi^{(2)}, \psi]^* W[\varphi^{(1)}, \chi] ,
  \end{equation*}
which is nothing but a simple identity about determinants (recall
$W[\varphi^{(1)}, \varphi^{(2)}] = 1$), we f\/ind that the surface terms can be
expressed in the desired form in terms of f\/inite quantities so these
limit numbers can be used for the purposes of $\Gamma^{}_1$,
$\Gamma^{}_2$. The examples coming in the following sections will show
this in close detail.

Now we can see the big practical advantage of the boundary value space
method to using von Neumann's characterization directly. Namely, here we
need to check square integrability of eigenfunctions only in a local
neighbourhood of the singularity while, in the von Neumann approach, we need
to do it over the whole conf\/iguration space. Besides, there are some
additional benef\/its as well.

First, in the von Neumann method, we need to f\/ind eigenfunctions for a
nonreal eigenvalue, while here for a real one. For example, many
eigenequations become simpler for the eigenvalue $E_0 = 0$.

Second, we actually need only one solution, as the condition $W[\varphi^{(1)},
\varphi^{(2)}] = 1$ allows us to f\/ind another one in a form
  \begin{equation*}
  \varphi^{(2)}(x) := \varphi^{(1)}(x) \int_{x_0}^x
  \frac{\d x}{\left[\varphi^{(1)}(x)\right]^2} .
  \end{equation*}

Third, we need the reference modes only approximately, to only such a
preciseness that the limit of their Wronskian with any $\psi$ can be
determined, plus that their local square integrability can be checked. In
practice, this usually means that we need to know their leading
and f\/irst subleading asymptotic behaviour.

In case our singularity is a so-called regular singular point then the
asymptotic solutions are in fact known from the Frobenius method. It is
usually advantageous to choose the regular solution for one of the
reference modes.

As a last virtue to mention, it is also apparent that the need for a
boundary condition, as well as its form, is decided only locally around
the singularity and not globally along the whole conf\/iguration space.
Using the von Neumann approach with the def\/iciency eigenfunctions this point
would also remain hidden. Actually, if we are on an interval bordered by
two limit-circle singular endpoints then we are allowed to use dif\/ferent
reference modes at the two endpoints.

Concerning the ef\/f\/iciency condition not yet discussed, the mathematical
literature ensures that this will also be fulf\/illed with the
$\Gamma^{}_1$, $\Gamma^{}_2$ chosen above. Indeed, it can be shown that,
to any two complex numbers $c^{(1)}$, $c^{(2)}$, one can f\/ind a $\psi$ in
the adjoint domain whose limit numbers are these $c^{(1)}$ and $c^{(2)}$.
Observing that a regular endpoint (where the potential does not diverge)
behaves the same way as a limit-circle singular endpoint from all
relevant aspects -- we can use $\chi^{(1)}(x) = -x$ and $\chi^{(2)}(x) =
1$ as approximate reference modes -- one can obtain that, in the special
case of a regular endpoint, the limits of $\psi$ and $\psi'$ as boundary
values do satisfy the ef\/f\/iciency condition.

The presented description of the self-adjoint domains via reference modes
contains some arbitrariness. One such freedom is in the value of
$E_0$. Since the term containing $E_0$ in the eigenequation is
overwhelmed by the diverging potential term, it is plausible to expect
and can actually be proven that the asymptotic behaviour of the reference
modes that decides the limit numbers of the wave functions does not
depend on $E_0$. The other arbitrariness is how the two independent
reference modes are chosen for a given $E_0$. This latter uncertainty,
which is an $SL (2,\R)$ amount of freedom, really inf\/luences the
characterization. (As a special case of what has been mentioned about the
non-uniqueness of $\Gamma$s in Section~\ref{sec:math}.) The family of
self-adjoint domains is nevertheless the
same, what changes is only that which domain is indexed by which~$U$.
Correspondingly, parameters chosen to parametrize $U \in {\cal U}(2)$ may
also change. Naturally, it is advantageous if we are able to introduce
parameters in such a -- reference mode-dependent -- way that they remain
the same under changing the reference modes.

A last remark is that, should our Hamiltonian have a kinetic term with
$\frac{\d}{\d x} \big( p(x) \frac{\d\psi}{\d x} \big)$ instead of
$\frac{\d^2\psi}{\d x^2}$, we only need to replace $\frac{\d\psi}{\d x}$
with $p(x) \frac{\d\psi}{\d x}$ in the above formulas and considerations.
Any possible zeros of $p(x)$ are also to be considered singularities
which, with this replacement, can be treated analogously to the
singularities of an operator with constant $p$. An example is provided by
the operator $ - \frac{\hbar^2}{2m} \left[ \frac{\d}{\d x}
\left( x^4 \frac{\d}{\d x} \right) + 2 x^2 \right] $ mentioned above.
The zero of $p(x) = x^4$, $x=0$, indicates a singular point there. The
reference modes are to be normalized to $ p W[\varphi^{(1)}, \varphi^{(2)}] = 1$,
and can be chosen for $E_0 = 0$ as $\varphi^{(1)}(x) = x^{-1}$,
$\varphi^{(2)}(x) = - x^{-2}$. These reference modes indeed behave singularly
at $x = 0$, and show that, in this case, the singularity is of limit-point
type as neither of them is square integrable in a f\/inite neighbourhood of
$x=0$.

\section{Singular potential on a half line}  \label{sec:halfsing}

Let us again take two examples  --  actually, the generalizations of our
previous two examples -- to apply the knowledge collected in the
previous section.

\setlength{\figwidth}{.3\textwidth}
\setlength{\txtwidth}{\textwidth} \addtolength{\txtwidth}{-1.3\figwidth}
\miniabove
\begin{minipage}[b]{\txtwidth}
\miniparindent
First, let us consider a half line again, but now with a potential that
diverges in a limit-circle way at the f\/inite endpoint.
Based on the f\/indings and notations introduced in
Sections~\ref{sec:halffree} and \ref{sec:singpot}, we do not need much
explanation why the
boundary value space approach and the corresponding
\minijustify
\end{minipage}
\hfill
\begin{minipage}[b]{\figwidth}
\raisebox{.05\textwidth}{\resizebox{\textwidth}{!}
{\includegraphics{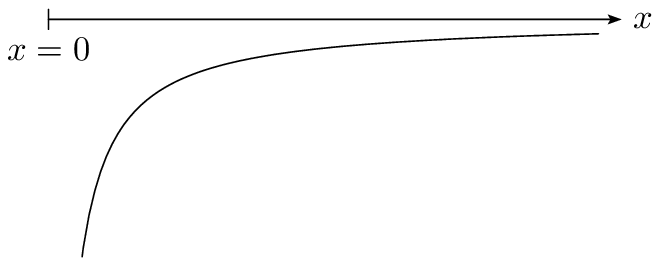}}}
\end{minipage}
\minibelow
characterization of self-adjoint domains can be given as\footnote{An $L_0$
may be needed to introduce here or there, because of the already
mentioned dimensional reason, but it is also possible that a coef\/f\/icient
from the potential can take this role. $L$ in \eqref{eq:aau} may also have a
dimension dif\/ferent than length, depending on the conventions we choose.}
  \begin{equation}  \label{eq:aau}
  {\cal H}_{\rm b} = \C ,  \qquad
  \Gamma^{}_1 \psi = W[\varphi^{(1)}, \psi](+0) ,  \qquad
  \Gamma^{}_2 \psi = W[\varphi^{(2)}, \psi](+0) ;  \qquad
  \Gamma^{}_1 \psi + L \Gamma^{}_2 \psi = 0 .
  \end{equation}
To be concrete, we may choose the three dimensional Coulomb problem, with
the Hamiltonian
  \begin{equation*}
  H = \frac{\hbar^2}{2m} \left( - \bigtriangleup + \frac{g}{r} \right) .
  \end{equation*}
After the usual separation of variables in spherical coordinates
(including the radial mapping $ L^2 \left( (0, \infty), r^2 \d r \right)
\to L^2 \left( (0, \infty), \d r \right) $), the radial part,
  \begin{equation*}
  H_{\rm rad}^{(l)} \psi^{\vphantom{|}}_{\rm rad} (r) =
  \frac{\hbar^2}{2m} \left( - \frac{\d^2}{\d r^2} + \frac{l(l+1)}{r^2}
  + \frac{g}{r} \right) \psi^{\vphantom{|}}_{\rm rad} (r)
  \end{equation*}
will have an ambiguity in self-adjointness in the $l=0$ angular momentum
channel because the centre is a limit-point singularity for the radial
Hamiltonians $H_{\rm rad}^{(l)}$ for $l > 0$ but is a limit-circle one
for $l = 0$.

For $l=0$, a satisfactory approximation for reference modes near
$r = 0$ is
  \begin{equation}  \label{eq:aax}
  \varphi^{(1)}(r) \approx -r , \qquad
  \varphi^{(2)}(r) \approx 1 + gr {\rm ln} |g|r .
  \end{equation}
These can be used in the boundary condition \eqref{eq:aau}. For
$L = 0$, the widely known and usually considered case yields, in which
the $\psi$s allowed by the boundary condition are regular at the
centre. In the other cases $L \ne 0$, a singular component is
present in $\psi$ so $\psi'$ diverges at the centre. Experiments make us
quite conf\/ident in that, when the potential is intended to express a~purely
electromagnetic interaction, we should choose $L=0$ in addition. However,
for a physical situation where some additional, short-range, interaction
is expected to be present between the centre and the particle, a model
with $L \ne 0$ is a good candidate. Mesic atoms, in which not electrons
but some negatively charged mesons like $\pi^-$ move around the
proton-containing centre, provide an example for this, because of the
short-range strong force also acting between the protons and mesons.

\setlength{\figwidth}{.32\textwidth}
\setlength{\txtwidth}{\textwidth} \addtolength{\txtwidth}{-1.3\figwidth}
\miniabove
\begin{minipage}[b]{\figwidth}
\raisebox{.05\textwidth}{\resizebox{\textwidth}{!}
{\includegraphics{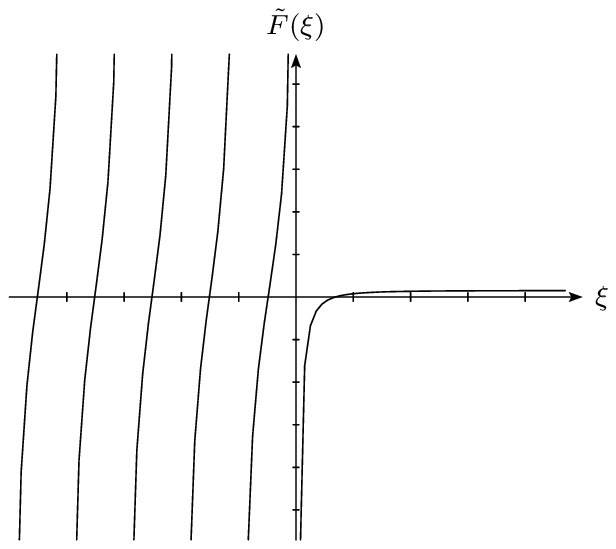}}}
\end{minipage}
\hfill
\begin{minipage}[b]{\txtwidth}
\miniparindent
For generic $L$, the condition for bound states can be found
\cite{ref:AGHH} to be
  \begin{equation}  \label{eq:aay}
  g \tilde{F}\left(\frac{g}{2\sqrt{-2mE/\hbar^2}}\right) = -\frac{1}{L}
  \end{equation}
with
  \begin{equation*}
  \tilde{F}(\xi) = \Psi(1+\xi) - \ln |\xi| -
  \frac{1}{2\xi} - \Psi(1) - \Psi(2) ,
  \end{equation*}
where $\Psi$ denotes the di-Gamma function.
\end{minipage}

If our potential is attractive, $g < 0$, then, for $L = 0$, we obtain the
well-known bound state energies
  \begin{equation*}
  E_n = - \frac{R}{n^2} ,  \qquad  n = 1, 2, \ldots  \qquad  (L = 0) .
  \end{equation*}
However, taking for example the case $L = \infty$, the solutions of the
transcendental equation~\eqref{eq:aay} give a dif\/ferent inf\/inite sequence
of bound states in the $l=0$ angular momentum channel:
  \begin{equation*}
   E_n = - \frac{R}{(n-c_n)^2} ,  \qquad  n = 1, 2, \ldots  \qquad
  (L = \infty)
  \end{equation*}
with
  \begin{equation*}
  c_1 \approx 0.5130 ,  \qquad  c_2 \approx 0.4879 ,  \qquad
  c_3 \approx 0.4857 ,  \qquad  \ldots,  \qquad  c_\infty \approx 0.4844
  \;\; (\hbox{limit}).
  \end{equation*}

The scattering properties -- like the scattering length and phase shift
quant\-ities --  and other physical aspects are similarly considerably
inf\/luenced by the value of the self-adjointness para\-meter $L$.

\section{Singular potential on a line}  \label{sec:linesing}

\setlength{\figwidth}{.30\textwidth}
\setlength{\txtwidth}{\textwidth} \addtolength{\txtwidth}{-1.3\figwidth}
\miniabove
\begin{minipage}[b]{\txtwidth}
Next, let us again extend our conf\/iguration space to a line, on which the
potential admits a singularity that is of limit-circle type from both
directions. Similarly to the ``free line with a pointlike singular
object" example seen in Section~\ref{sec:halffree}, the two sides of the
singularity double the dimension of the
\minijustify
\end{minipage}
\hfill
\begin{minipage}[b]{\figwidth}
\raisebox{.18\textwidth}{\resizebox{\textwidth}{!}
{\includegraphics{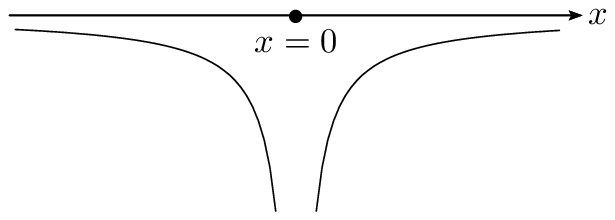}}}
\end{minipage}
\minibelow
boundary value space, and we also need to double (to extend) the
reference modes. Having seen~\eqref{eq:aaq} and~\eqref{eq:aau}, no wonder
that we choose
  \begin{equation*}
  {\cal H}_{\rm b} = \C^2 ,  \qquad
  \Gamma^{}_1 \psi = \begin{pmatrix} W[\varphi^{(1)}, \psi](+0) \cr
  W[\varphi^{(1)}, \psi](-0) \end{pmatrix} ,  \qquad
  \Gamma^{}_2 \psi = \begin{pmatrix} W[\varphi^{(2)}, \psi](+0) \cr
  -W[\varphi^{(2)}, \psi](-0) \end{pmatrix} ,
  \end{equation*}
and obtain the boundary conditions, with $U \in {\cal U}(2)$, as
  \begin{equation*}
  \big( U - \mathbf{1}^{}_{\C_{}^2} \big) \Gamma^{}_1 \psi +
  i \big( U + \mathbf{1}^{}_{\C_{}^2} \big) \Gamma^{}_2 \psi = 0 .
  \end{equation*}
Here again, a certain subfamily of the family of self-adjoint
Hamiltonians contains the separated cases where the system decouples to
two independent half line subsystems. This means that, in those cases,
the particle bounces back from the singularity. This phenomenon is
present for attractive potentials as well, which is surprising for the
physical intuition. It creates a~feeling that, in those cases, something
repulsive must happen at the location where the potential diverges.

\setlength{\figwidth}{.30\textwidth}
\setlength{\txtwidth}{\textwidth} \addtolength{\txtwidth}{-1.3\figwidth}
\miniabove
\begin{minipage}[b]{\txtwidth}
\miniparindent
On the other side, in the nonseparated cases we can observe
ref\/lection-transition, time delay and f\/ilter phenomena similarly to the
case of a pointlike interaction on a line. This includes that we f\/ind
tunneling through a repulsive potential as well
\cite{ref:DE}, which is much more
surprising
\minijustify
\end{minipage}
\hfill
\begin{minipage}[b]{\figwidth}
\raisebox{.15\textwidth}{\resizebox{\textwidth}{!}
{\includegraphics{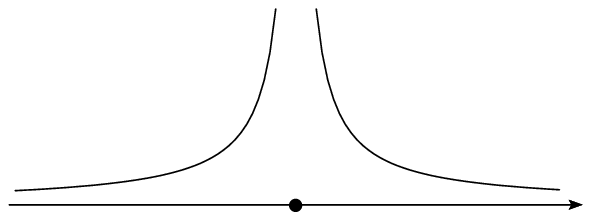}}}
\end{minipage}
\minibelow
for such a ``fat" inf\/inite potential barrier than for a ``thin" pointlike
object. Again, one may try to illustrate these cases physically by that
some attractive ef\/fect is present at the location where the potential is
undef\/ined.

Many other physical properties, like integrability for Calogero-type
potentials whose singular term $\frac{g}{r^2}$ has
$g < \frac{3\hbar^2}{8m}$ \cite{ref:Calpaper}, are also dependent on the
self-adjoint version chosen\footnote{Writing $\frac{g}{r^2}$ as
$\frac{\hbar^2}{2m} \frac{l(l+1)}{r^2}$, ambiguity in the
self-adjointness occurs for $l < \frac{1}{2} $. In parallel, the
Hamiltonian will be bounded from below when $g \ge - \frac{\hbar^2}{8m}$
(when $l$ is real).}. These dependences
might be utilized in the future to build quantum devices like quantum
gates, f\/ilters and qubits when the boundary parameters become tunable
and controllable \cite{ref:soroban}.

\section{Conclusion}  \label{sec:concl}

We have seen how the ambiguity in the self-adjointness of a Schr\"odinger
Hamiltonian with a~singular potential can be understood, and how it can
be technically treated in a framework that requires only a reduced amount
of calculational ef\/forts.

Naturally, what has been presented here is only a narrow
and practice-oriented extract from the extensive research f\/ield of
boundary conditions and singular potentials and, more generally, of
self-adjoint extensions. For further reading, in addition to the works
already referred to, one can start with consulting, for example,
\cite{ref:GTV,ref:FLS,ref:Krall,ref:Esteve,ref:CH,ref:EG}
and references therein.

To close the discussion with the message: the self-adjointness ambiguity
caused by irregulari\-ties of the potential can be interpreted such that
the system carries extra physical properties which can not be expressed
via the potential function but through a boundary condition at each
irregularity. These properties are to be f\/ixed either by some additional
theoretical knowledge or physical information about the system, or by
measurement, f\/itting the unknown boundary parameters to experimental
data.

\subsection*{Acknowledgements}

Work supported in part by the Czech Ministry of Education, Youth and Sports
within the project LC06002.

\pdfbookmark[1]{References}{ref}
\LastPageEnding

\end{document}